# An Evaluation of Lightweight Deep Learning Techniques in Medical Imaging for High Precision COVID-19 Diagnostics

[1]Ogechukwu Ukwandu
ukwanduo@cardiff.ac.uk
Department of Computer, Communication and Information Systems, School of Engineering and Built Environment, Glasgow Caledonian University & School of Computer Science and Informatics, Cardiff University.

[2]Hanan Hindy
hanan.hindy@cis.asu.edu.eg
Computer Science Department, Faculty of Computer and Information Sciences, Ain Shams University, Cairo 11566, Egypt

[3]Elochukwu Ukwandu
eaukwandu@cardiffmet.ac.uk
Department of Applied Computing and Engineering, Cardiff School of Technologies, Cardiff Metropolitan University, Cardiff CF5 2YB, UK

**Correspondent author:** Elochukwu **Ukwandu,** eaukwandu@cardiffmet.ac.uk
Department of Applied Computing and Engineering, Cardiff School of Technologies, Cardiff Metropolitan University, Cardiff CF5 2YB, UK.

## ABSTRACT

Timely and rapid diagnoses are core to informing on optimum interventions that curb the spread of COVID-19. The use of medical images such as chest X-rays and CTs has been advocated to supplement the Reverse-Transcription Polymerase Chain Reaction (RT-PCR) test, which in turn has stimulated the application of deep learning techniques in the development of automated systems for the detection of infections. Decision support systems relax the challenges inherent to the physical examination of images, which is both time consuming and requires interpretation by highly trained clinicians. A review of relevant reported studies to date shows that most deep learning algorithms utilised approaches are not amenable to implementation on resource-constrained devices. Given the rate of infections is increasing, rapid, trusted diagnoses are a central tool in the management of the spread, mandating a need for a low-cost and mobile point-of-care detection systems, especially for middle- and low-income nations. The paper presents the development and evaluation of the performance of lightweight deep learning technique for the detection of COVID-19 using the MobileNetV2 model. Results demonstrate that the performance of the lightweight deep learning model is competitive with respect to heavyweight models but delivers a significant increase in the efficiency of deployment, notably in the lowering of the cost and memory requirements of computing resources.

**Keywords:** Diagnostic analytics; COVID-19 detection; lightweight deep learning techniques; point-of-care; resource-constrained devices; machine learning.

## 1.0 Introduction

Medical imaging captures a visual representation of internal organs of the body to facilitate the detection and in turn, the diagnosis of health conditions as well as a means to confirming patient responses to treatments [1]. The use of X-rays, Computed Tomography (CT), Magnetic Resonance Imaging (MRI), Positron Emission Tomography (PET), and Ultrasound imaging [2] is well-established, a core enabler for clinicians to determine appropriate diagnoses and reach timely decisions on optimum treatments.

A continual increase in the use of and reliance on images has been evident over the recent past [3], with concomitant ramifications through increases to the workloads of clinicians. Poor (high) clinician to patient ratios, especially the case in low and middle-income countries, has motivated the development of high precision, intelligent based automated systems that can provide decision support in interpreting medical images for widespread appropriate and timely diagnoses. The ongoing COVID-19 pandemic, which has created public health crises worldwide resulting in ~505 million cases [4] and 6.2 million deaths as of April 2022 [5], has intensified the need for such systems. The use of chest X-rays and CT images has been advocated as a complementary test for the early detection and accurate diagnosis of COVID-19 [6], [7], [8], [9].

Currently, one of the principal means of testing for COVID-19 is the Reverse-Transcription Polymerase Chain Reaction (RT-PCR), which detects viral nucleic acids [7] with 63% positivity [9] using nasal swabs. The technique presents limitations, such as availability due to high demand and a significant false-negative rate [6], resulting in incorrect diagnoses particularly in the early stage of infection. In the same vein, several hours or even days are required to receive the result. The medical community has proposed the use of chest X-ray and CT images as complementary tests in support of the early detection of COVID-19 as the virus causes infections in the lungs, as well as being useful in the monitoring of the rapid progression of the disease from early to critical stage. The latter is based on the observation of the level of irregular ground-glass opacities, consolidation, and pleural fluid mostly in the lower lobes, in the early stage and pulmonary consolidation at the critical stage [6] [8]. Chest X-ray images are most commonly used in COVID-19 detection due to widespread availability of equipment, affordability and the lower radiation dose applied compared to computed tomography (CT) [10]. Traditionally, medical images are at the heart of a physical examination and are subject to human interpretations. The interpretation of these images in tandem with the accompanying patient medical records is time consuming and requires execution by appropriate clinical experts. A deep understanding of internal organ structures and careful observation is a necessity or partially accurate diagnoses, poor treatment regimens and management of diseases, especially in developing countries, result as a consequence. Thus there is an unambiguous need for an accurate system supports the interpretation and analysis of chest X-ray images for early detection of COVID-19 [11].

Here, a lightweight deep learning Convolutional Neural Network (CNN)-based MobileNetV2 model is proposed for the detection of COVID-19 using chest X-ray images as the input. A critical review of the body of research reporting on the development of heavyweight deep learning CNN models in detecting COVID-19 using chest X-ray images is provided, principally to establish a robust reference with which the performance of the proposed approach is compared. The results obtained are compared using performance metrics such as accuracy, precision and sensitivity. The dataset used in the development of the proposed model is the publicly available COVID-19 dataset from Kaggle [12].

The remainder of the paper is organised as follows: Section 2 is a review of relevant reported literature, the aim being to provide a benchmark of performance for comparison with the proposed approach and to surface the gap in knowledge that motivated this research; Section 3 details the materials and methods used in this research; Section 4 presents an evaluation of results; Section 5 discusses the potential of the approach; and conclusions are drawn in Section 6 with recommendations for future research.

## 2.0 State-of-the-Art

The review of reported literature on existing and emerging trends in the use of deep learning techniques in medical diagnosis using medical imaging has been segmented by the most common organ-specific areas of Breast, Brain, and Lungs.

The earliest publication [13] reporting the use of deep learning in medical imaging detailed the detection of micro-calcifications for mammography appeared in 1994, in which Zhang *et al.* [14] applied a shift-invariant network to eliminate the degree of false-positives in a computer-aided (CADx) scheme; individual micro-calcifications within a Region of Interest (RoI) as reported by the CAD system were identified. Results using 34 mammograms and 168 RoIs showed that the network achieved an average area under the Receiver Operating Curve (ROC) of 0.91 and eliminated 55% of the false-positive without loss of the true-positive performance. Similar research was reported by Yap *et al.* [15], Al-antari *et al.* [16], Brya *et al.* [17], and Fujioka *et al.* [18].

Ismael *et al.* [19] proposed a model for the classification of brain tumours implemented through the ResNet50 architecture using 3064 MRI images capturing 3 brain tumour types - Meningiomas, Gliomas, and Pituitary. The data were split into 80% for training and 20% for test. Result showed that the proposed model yielded an accuracy of 99%, out-performing previously reported algorithms - based on shallower architectures than ResNet50that - developed using the same dataset. The main guidance arising from the research was that a network with sufficient depth, such as ResNet50, was required to achieve an acceptable performance due to the diversity and complexity of medical images; however, this conclusion requires more extensive evaluation using larger tumour datasets. Rehman *et al.* [20] and Sarhan [21] also demonstrated the feasibility of the application of deep learning in the detection of brain tumours using MRI images. A human-machine collaborative design strategy based on a residual architecture methodology, entitled 'COVID-Net', was reported by Wang *et al.* [22]; a deep CNN for the detection of COVID-19 using chest X-rays demonstrated the potential of deep learning in lung medical imaging. Other, related research that utilised lightweight deep learning techniques is detailed in Abraham and Nair [23], Wang *et al.* [24], and Apostolopoulos *et al.* [25]. Moreover, Hall *et al.*,[26] described the development of a ResNet 50-based model which provided an accuracy of 89.2% and the Maghdid *et al.* [27] pre-trained AlexNet model yielded an accuracy of 98%. [46] reported on two lightweight deep-CNN models - MobileNetV2 and SqueezeNet – that gave an accuracy of 84.3%, and 73.2%, respectively; however, both models were implemented using a small dataset and therefore a more robust validation of the reported results on larger dataset remains. 'CoroNet', a deep CNN based on Xception architecture [28] achieved an overall accuracy of 89.6%. In [29], 'COVIDiagnosis-Net', based on deep CNN lightweight SqueezeNet with Bayesian optimisation was detailed, the model achieving an overall accuracy of 76% without, and an accuracy of 98% with data augmentation, respectively. Ozturk, *et al.* [30], Mukherjee *et al.* [31] and Alzubaidi *et al.* [11] conducted reviews on the role of deep learning for the early detection of COVID-19. The reviews concluded that the datasets used for both training and test were imbalanced, principally due to absence of COVID-19 chest X-ray images.

Zouch *et al.* [32] opined that the early detection of COVID-19 related anomalies remained the major challenge to be overcome in arresting the spread of the virus, with the application of AI being one of the approaches with the potential to detect the features of relevance that characterise COVID-19 as well as being integral to creating effective solutions. VGG and ResNet deep learning models that used chest X-ray images only were explored, yielding an accuracy of 99.35% and 96.77% respectively.

Bhattacharyya, *et al*. [33] proposed a three-step process of the segmentation of X-ray images using the conditional generative adversarial network (C-GAN) to analyse the lung. The segmented lung images are then fed into a novel pipeline that combines key points extraction methods and trained deep neural networks (DNN)to extract the relevant discriminatory features. An accuracy of 96.6% using the VGG-19 model with the binary robust invariant scalable key-points (BRISK) algorithm was claimed.

Koh *et al.* [34] argued that deep learning techniques could improve the sensitivity of the detection of incidental breast cancers from chest computed tomography (CT) scans. 'RetinaNet', a deep learning algorithm based on chest CTs was developed to detect breast cancer and results showed that the sensitivity of detection was between 88.5% - 96.5% for internal test sets and 90.7% and 96.1% for external test sets. The conclusion was that sufficient evidence was provided that proved the feasibility of deep learning algorithm-based approaches could detect breast cancer on chest CT in both the internal and external test sets at a meaningful sensitivity.

Other approaches that have reported the use and evaluated the performance of deep learning algorithms in the diagnosis of COVID-19 applications with X-ray images include the use of space transformer network (STN) with CNNs by Soni *et al.* [35], Alqudah, Qazan and Alqudah [36], Chakraborty, Dhavale and Ingole [37], Karakanis and Leontidis [38], Bekhet *et al*. [39]; Huang and Liao [40] developed lightweight CNNs using Chest X-ray images based COVID-19 detection. Zebin and Rezvy [41], Nayak et al., [42], Sanida [43], Heidari et al., [44], Asif, Zhao, Tang and Zhu [45], Kogilavani, Prabhu, and Sandhiya [46] and Rajawat, Hada, Meghawat, and Lalwani [47] report on similar approaches. Subramanian, Elharrouss, and Al-Maadeed [48] provided a review on deep learning-based detection methods for COVID-19.

## 2.1 Research Gap

Evident from the review of the extensive research reported recently in the development of AI systems founded on CNN for COVID-19 detection, that most of the proposed systems are based on heavyweight deep learning techniques that require substantial computing power and thus are not amenable to deployment on resource constrained devices. Thus, here, an evaluation of the feasibility of relying on lightweight deep learning (CNN) models using chest X-ray images amenable to implementation on resource constrained devices in COVID-19 diagnosis integrated in high-level precision decision support systems is presented. For clarity, an exemplar of a resource-constrained device is Smartphones, a platform that enjoys significant adoption and cheaper to access, particularly in low and middle-income countries.

## 3.0 Dataset

Kaggle [12], a publicly available COVID-19 dataset from created by a team of researchers from Qatar University and University of Dhaka in partnership with organisations in Pakistan and Malaysia and medical practitioners is used in the development. The extensive collaboration has generated diversity in the dataset, a characteristic that is central to the deep learning training process as it comprises of both patterns associated with the image acquisition process, as well as signatures of the disease; the dataset consists of chest X-ray images of COVID-19 positive cases together with Normal and Viral Pneumonia images. Kaggle comprises **1200** (30.88%) COVID-19 positive images, **1341 (34.51%)** normal images, and **1345** (34.61%) viral pneumonia images. All the images are in the Portable Network Graphics (PNG) file format with resolution of 1024-by-1024, and 256-by-256 pixels. The dataset continues to be updated; therefore, the number and scope of images is likely to increase in the future.

## 3.1 MobileNetV2 with Transfer Learning

The effective training of deep learning models requires large, diverse datasets comprising a significant number of samples and most often, ready access to high computational resources, the extent of the latter determining the time to execute the training phase. One of the major barriers in the training of deep learning models is that the datasets required are governed by strict levels of confidentiality in turn limiting availability, especially in the medical domain. Transfer learning, where the knowledge and weights, from a (pre-trained) model trained on a vast number of samples in a different, but related field is the foundation to train a new model subject to limited access to samples [49], has been used as a solution to insufficient datasets and restricted computing resources. The pre-trained model is fine-tuned by freezing the first layers which detect the common features across the application, with the deeper layers then re-purposed and trained to classify the categories within the target training dataset. Not only number of samples needed minimised but also a significant reduction in the length of time needed to train a new model is achieved. The proposed model employs the MobileNetV2 CNN architecture, pre-trained on ILSVRC-12 challenge ImageNet dataset [50] which contains millions of images with 1000 classes. MobileNetV2[51], a lightweight CNN network architecture with fewer parameters compared to other, heavyweight models such as VGGs and ResNets was chosen as as it can be trained efficiently and be deployed on low-storage, low-power devices such as mobile phones, thus reducing the barriers to widespread adoption by medical practitioners, especially in rural and low-income areas.

## 3.2 Developmental Methodology

The development of the model followed four stages; (a) the pre-processing of images (b) the training of a 3-class MobileNetV2 model to classify chest X-ray images with COVID-19, Normal, Viral Pneumonia infection and two models for a 2-class classification (COVID-19 and Normal) model and 2-class classification (COVID-19 and Viral Pneumonia); (c) an evaluation the performance of the model (d) a comparison of model performance with reported heavyweight deep learning CNN models in COVID-19 detection using chest X-ray images. A block diagram of the methodology is shown in Figure 1.

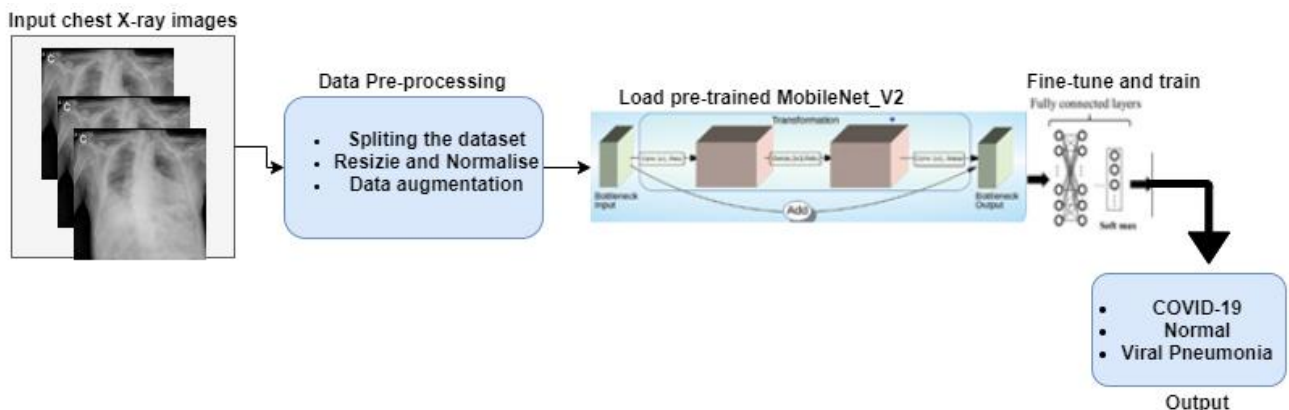

Figure 1: The four stages of the developmental methodology

### 3.2.1 Image pre-processing

In summary, the stages of the pre-processing of images are as follows:

- **Segmentation into training, validation, and test sets.**

  Images were split into training, validation, and test sets using the ratio 82.35%, 7.72% and 9.93%, respectively. The resulting number of images for each set for the 3-class model is shown in Table 1 and that of the 2-class models in Table 2 and Table 3.

Table 1: The distribution of the chest X-ray dataset for the 3-class model

| Training Set | | | Validation Set | | | Testing Set | | |
|---|---|---|---|---|---|---|---|---|
| Total Sample | | | Total Sample | | | Total Sample | | |
| 3200 (82.35%) | | | 300 (7.72%) | | | 386 (9.93%) | | |
| COVID | Norm | vPneumonia | COVID | Norm | vPneumonia | COVID | Norm | vPneumonia |
| 1000 | 1100 | 1100 | 100 | 100 | 100 | 100 | 141 | 145 |

**Legend**: vPneumonia = Viral Pneumonia, Norm = Normal, COVID = COVID-19

Table 2: The distribution of the dataset for 2-class (C0VID-19 and Normal) model

| Training Set | | Validation Set | | Testing Set | |
|---|---|---|---|---|---|
| Total Sample | | Total Sample | | Total Sample | |
| 2100 (82.64%) | | 200 (7.87%) | | 241 (9.48%) | |
| COVID-19 | Normal | COVID-19 | Normal | COVID-19 | Normal |
| 1000 | 1100 | 100 | 100 | 100 | 141 |

Table 3: The distribution of the dataset for 2-class (COVID-19 and vPneumonia)

| Training Set | | Validation Set | | Testing Set | |
|---|---|---|---|---|---|
| Total Sample | | Total Sample | | Total Sample | |
| 2100 (82.51%) | | 200 (7.86%) | | 245 (9.63%) | |
| COVID-19 | Viral Pneumonia | COVID-19 | Viral Pneumonia | COVID-19 | Viral Pneumonia |
| 1000 | 1100 | 100 | 100 | 100 | 145 |

- **Resizing and normalisation**

  The images were then resized to 224 x 224 pixels, the default size for in training deep learning models the MobileNetV2 classifier and transformed into a 3-channel grayscale; Thus, the final input image dimension is $224 \times 224 \times 3$. Normalisation of each of the image channels using mean = [0.0960, 0.0960, 0.0960] and standard deviation = [0.9341, 0.9341, 0.9341] followed.

- **Application of data augmentation**

  The data augmentation techniques applied were random sized crop and random rotation range of $15^o$ to the training set to enhance size and quality further followed by shuffling of the images. Data augmentation increases the scope of the training data by creating different repetitions of the data, often applied to prevent over-fitting by acting as a regulariser and facilitating model convergence. The data enhancement allows the model to generalise features through exposure to different variations of the image.

### 3.2.2 Training and Validation

#### 3.2.2.1 Experiments

All experiments were implemented using the Python 3.7.10 programming language on Google Colab. The model was trained using the PyTorch 1.8.1+cu101 framework on a Colab GPU interface, which allocated a 12GB NVIDIA Tesla K80 GPU. The pre-trained MobileNetv2 used to train a model for COVID-19 detection using chest X-ray images, was modified by the replacement of the number of output classes in the last fully connected linear block with the number of classes in the chest X-ray image dataset viz. 3 for the 3-class model and 2 for the two 2-class models. The modified model was then re-trained end-to-end to classify chest X-ray images over 3 experiments, each of which was performed using the following parameters: 30 epochs, learning rate scheduler with maximum learning rate of 0.0001, weight decay of 1e-4, and Adam optimiser. The images (training and validation sets) were placed in a data loader with batch size of 32 for loading into the model.

**Experiment 1: The training of a 3-class model, which classified the chest X-ray images into COVID-19, Normal and Viral Pneumonia**

The distribution of the dataset used is shown in Table 1. The model took 82.96 minutes to train excluding evaluation, the training accuracy started from 0.603 and reached 0.979 while the validation accuracy started from 0.84 and reached 0.995, as shown in Figure 2. The lowest training loss and validation loss achieved are 0.068 and 0.029 respectively, shown in Figure 3.

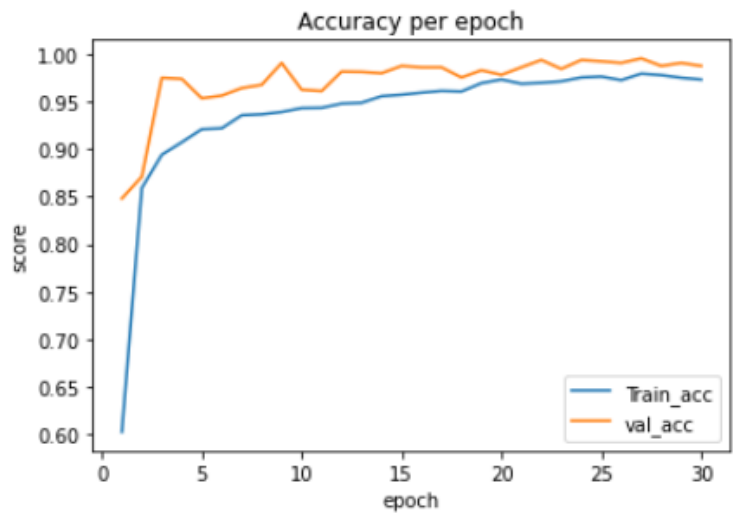

**Figure 2: 3-class model training and validation accuracy per epoch**

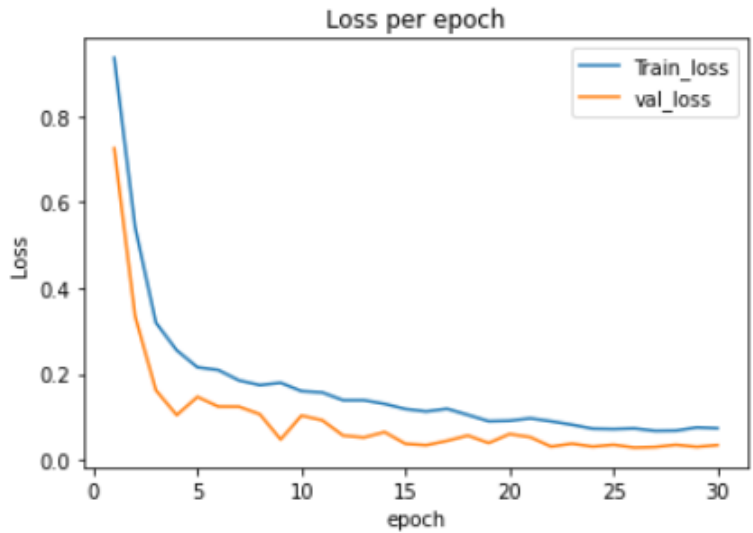

**Figure 3: 3-class model training and validation loss per epoch**

**Experiment 2**: **The training of a 2-class model which classified the chest X-ray images into COVID-19 and Normal.** The distribution of the dataset used is shown in Table 2. The model took 57.11 minutes to train excluding evaluation, the training accuracy started from 0.743 and reached 0.999 while the validation accuracy started from 0.754 and reached 1.000, as shown in Figure 4. The lowest training loss and validation loss achieved during training were 0.006 and 0.003 respectively, shown in Figure 5.

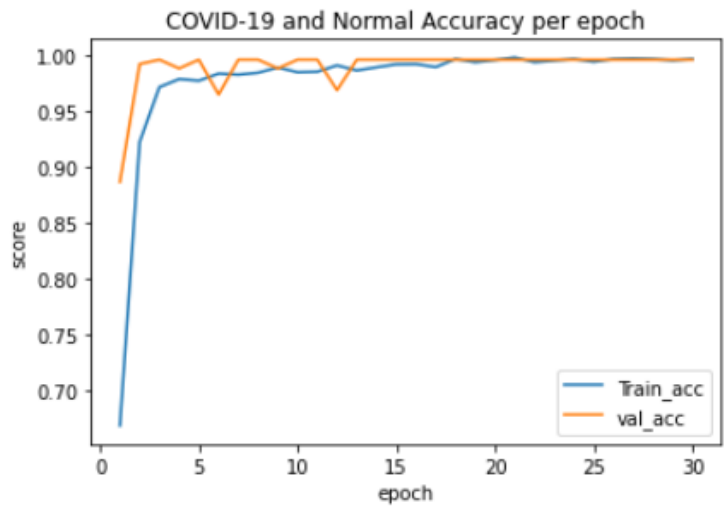

**Figure 4: COVID-19 and Normal training and validation accuracy per epoch**

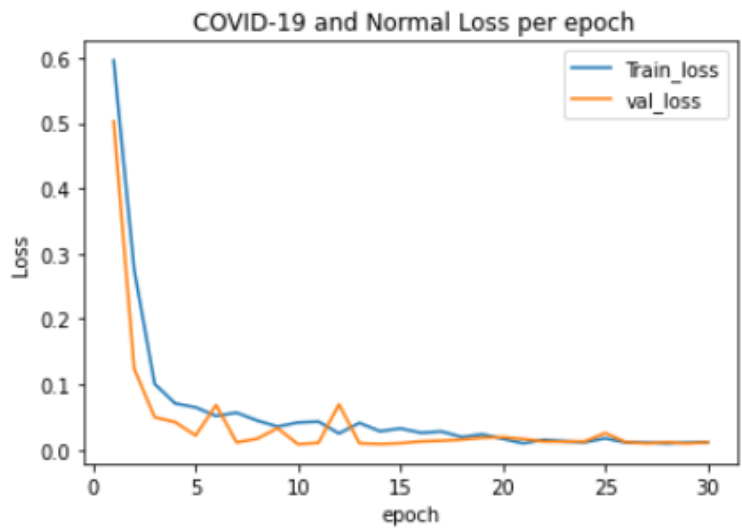

**Figure 5: COVID-19 and Normal training and validation loss per epoch**

**Experiment 3**: **The training of a 2-class model, which classified the chest X-ray images into COVID-19 and Viral Pneumonia.** The distribution of the dataset used is shown in Table 3. The model took 59.11 minutes to train excluding evaluation, the training accuracy started from 0.584 and reached 0.996 while the validation accuracy started from 0.785 and reached 1.000, as shown in Figure 6. The lowest training loss and validation loss achieved during training are 0.014 and 0.003 respectively, shown in Figure 7.

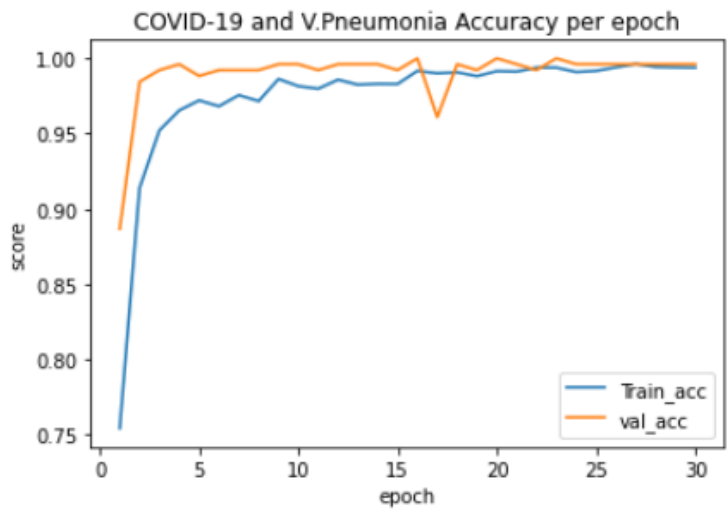

**Figure 6: COVID-19 and Viral Pneumonia training and validation accuracy per epoch**

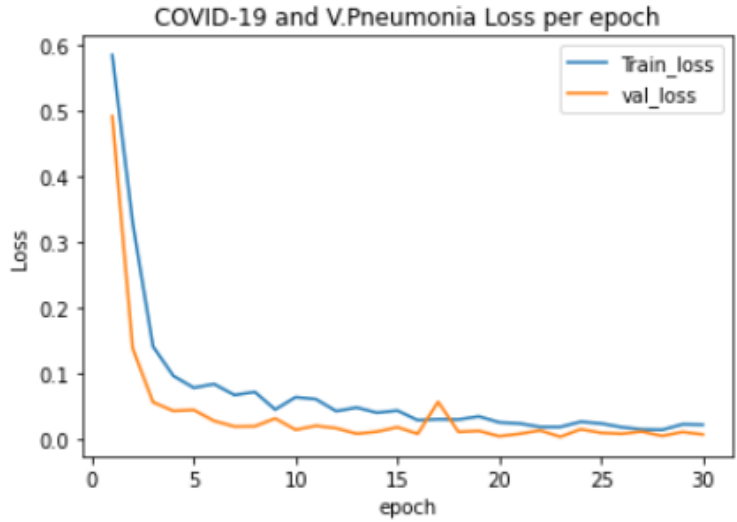

Figure 7: COVID-19 and Viral Pneumonia training and validation loss per epoch

### 3.2.2.2 Evaluation Metrics

Models were evaluated using statistical measures adopted routinely by the research community, such as accuracy, precision, recall, F-1 score, and mis-classification rate [52]. The measures were computed using the following confusion matrix parameters:

- True Positive (TP): the number of infected images classified correctly.
- True Negative (TN): the number of images correctly predicted as negative.
- False Positive (FP): The number of positive images incorrectly predicted as negative.
- False Negative (FN): The number of negative images incorrectly predicted as positive.

**Accuracy (ACC):** a measure of the percentage of cases correctly predicted out of the total number of cases calculated as:

$$ACC = \frac{TP+TN}{(TP+FP+FN+TN)} \quad (1)$$

**Precision (PRE):** a measure f the exactness of the model which is the percentage of the positive predicted cases that are true, calculated as:

$$PRE = \frac{TP}{(TP+FP)} \quad (2)$$

**Recall (REC):** a measure of the completeness of the model which is the percentage of positive case correctly identified to all the cases in a class, calculated as:

$$Recall = \frac{TP}{(TP+FN)} \quad (3)$$

**F1-Score:** captures both Precision and Recall in a single measure, especially significant for imbalanced dataset, calculated as:

$$F1 - Score = \frac{2 \; x \; precision \; x \; Recall}{(precision + Recall)} \quad (4)$$

**Mis-classification Rate (MIS):** a measure of the rate at which the model is wrong in predicting each class in the dataset, calculated as:

$$Mis - classification\ rate\ = \frac{FP+TN}{(TP+TN+FP+FN)} \qquad (5)$$

## 4.0 Evaluation of Results
### 4.1 3-Class Model
The results obtained for the 3-class model are summarised in Table 4; an overall accuracy of 94.5%, sensitivity of 92.3%, specificity of 96% and misclassification rate of 5% were achieved. The corresponding confusion matrix is shown in Figure 8.

Table 4: Evaluation of Results for the 3-Class Model

| Class | TP | TN | FN | FP | Acc (%) | Pre (%) | Recall (%) | Mis (%) | F-1 (%) | Sen (%) | Spe (%) |
|---|---|---|---|---|---|---|---|---|---|---|---|
| COVID-19 | 95 | 286 | 5 | 0 | 98 | 100 | 95 | 1.3 | 97 | 95 | 100 |
| Normal | 119 | 242 | 22 | 3 | 93 | 98 | 84 | 6 | 90 | 84 | 99 |
| Viral Pneumonia | 143 | 215 | 2 | 26 | 92 | 85 | 99 | 7 | 91 | 98 | 89 |

**Legend**: ACC = Accuracy, PRE = Precision, MIS = Misclassification Rate, F-1 = F-1 Score, SEN = Sensitivity, SPE = specificity.

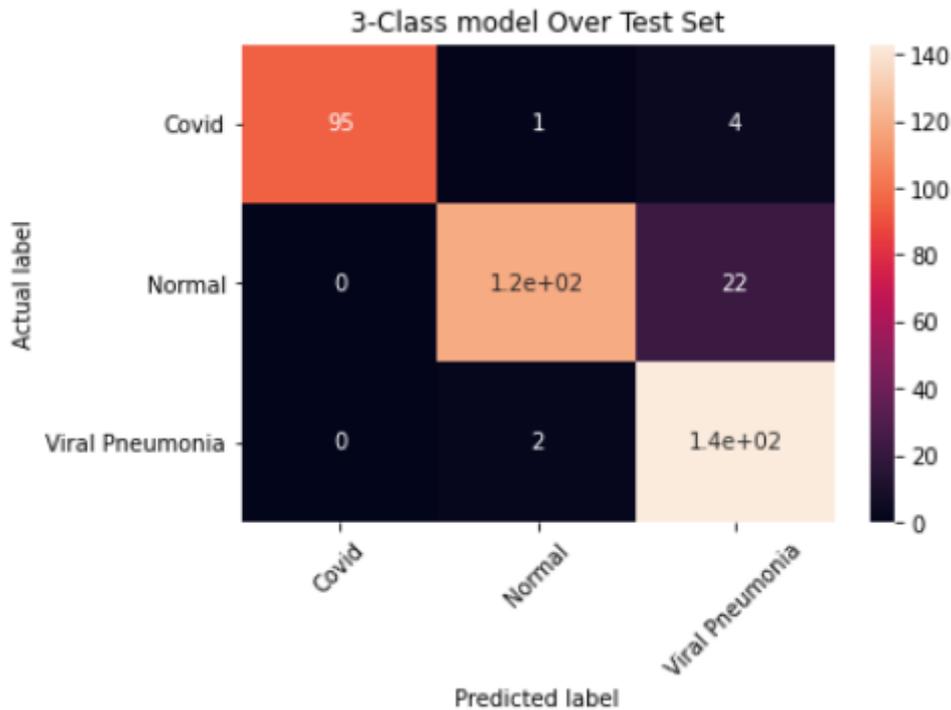

Figure 8: Confusion Matrix for the 3-Class Model

### 4.2 2-Class Model (COVID-19 and Normal)
The results obtained for the 2-class model (COVID-19 and Normal) are shown in Table 5; an overall accuracy of 99.6%, sensitivity of 99.5%, specificity of 99.5% and misclassification rate of 0.4% were achieved. The corresponding confusion matrix is shown in Figure 9.

**Table 5: Evaluation of Results for the 2-Class (COVID-19 and Normal) Model**

| Class | TP | TN | FN | FP | Acc (%) | Pre (%) | Recall (%) | Mis (%) | F-1 (%) | Sen (%) | Spe (%) |
|---|---|---|---|---|---|---|---|---|---|---|---|
| COVID-19 | 99 | 141 | 1 | 0 | 99.6 | 100 | 99 | 0.4 | 99 | 99 | 100 |
| Normal | 141 | 99 | 0 | 1 | 99.6 | 99 | 100 | 0.4 | 100 | 100 | 99 |

**Legend**: ACC = Accuracy, PRE = Precision, MIS = Misclassification Rate, F-1 = F-1 Score, SEN = Sensitivity, SPE= specificity.

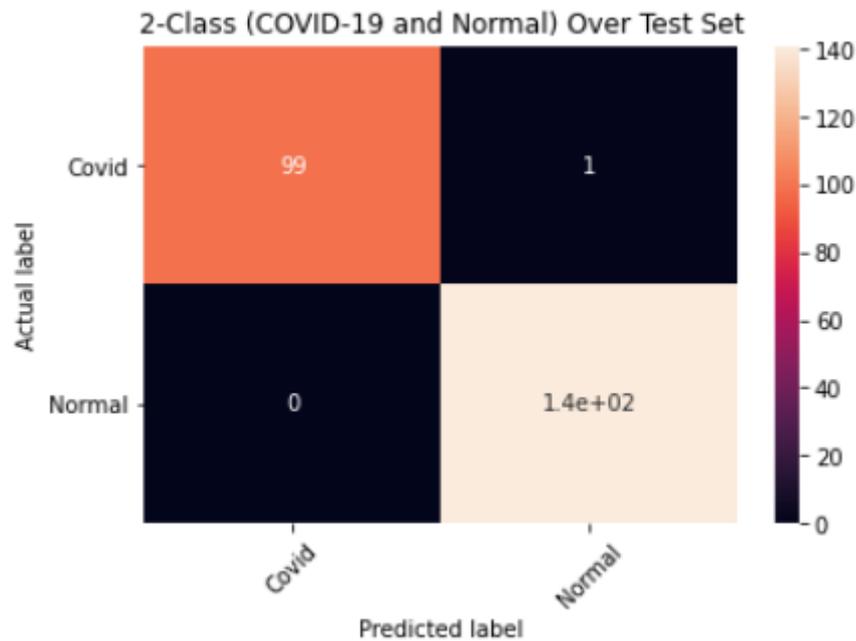

**Figure 9: Confusion Matrix for 2-Class Model (COVID-19 and Normal)**

## 4.3 2-Class Model (COVID-19 and Viral Pneumonia)

The results obtained for the 2-Class Model (COVID-19 and Viral Pneumonia) are shown in Table 6; an overall accuracy of 98.4%, sensitivity 98%, specificity of 98% and misclassification rate of 1.6% were achieved. The corresponding confusion matrix is shown in Figure 10.

**Table 6: Evaluation Result for the 2-Class (COVID-19 and V. Pneumonia) Model**

| Class | TP | TN | FN | FP | Acc (%) | Pre (%) | Recall (%) | Mis (%) | F-1 (%) | Sen (%) | Spe (%) |
|---|---|---|---|---|---|---|---|---|---|---|---|
| COVID-19 | 96 | 145 | 4 | 0 | 98.4 | 100 | 96 | 1.6 | 98 | 96 | 100 |
| Viral Pneumonia | 145 | 96 | 0 | 4 | 98.4 | 97 | 100 | 1.6 | 99 | 100 | 96 |

**Legend**:

**Legend**: ACC = Accuracy, PRE = Precision, MIS = Misclassification Rate, F-1 = F-1 Score, SEN = Sensitivity, SPE= specificity

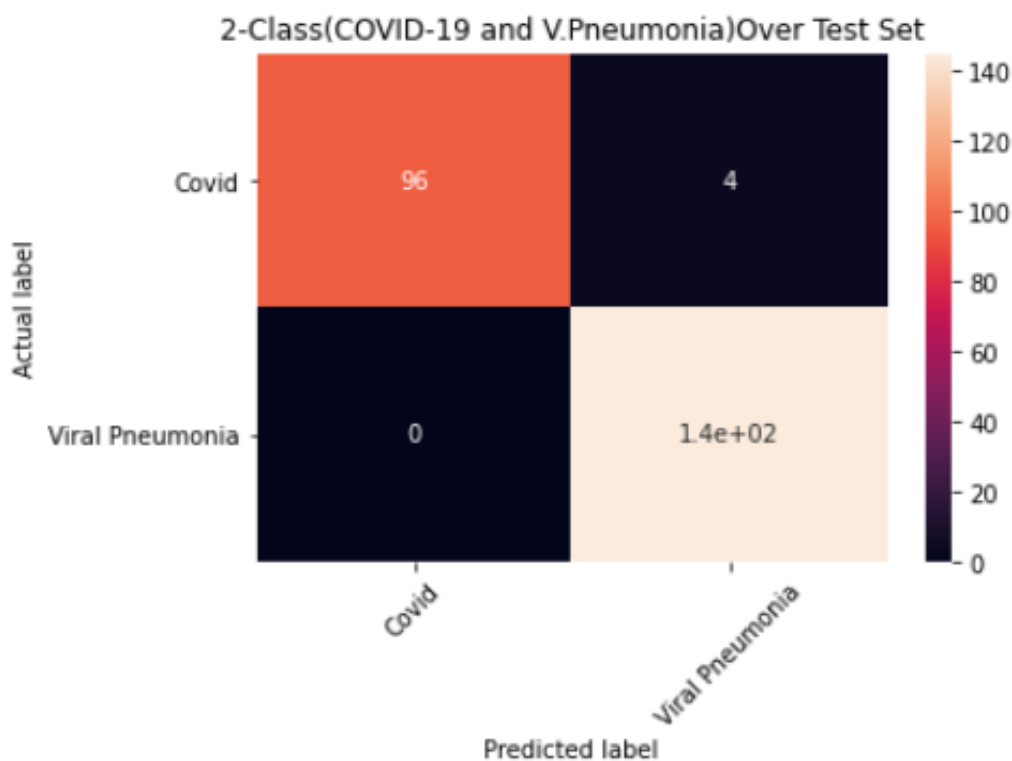

**Figure 1: Confusion Matrix for 2-Class Model (COVID-19 and V. Pneumonia)**

## 5.0 Discussions

An evaluation of the performance of lightweight deep-CNN architectures for efficient and rapid COVID-19 detection using chest X-ray images has been carried out. MobilenetV2, a lightweight deep-CNN architecture was adopted and trained with transfer learning using an extensive chest X-ray image dataset comprising 1200 COVID-19, 1341 Normal and 1345 Viral Pneumonia images to create an accurate model amenable to deployment on devices with low computational power such as mobile phones. Three models were trained viz. a 3-class model for the classification of COVID-19, Normal and Viral Pneumonia and two 2-class models for the classification of COVID-19, Normal and COVID-19, Viral Pneumonia.

The 3-class model, presented in Table 4, provided an overall accuracy of 94.5%. The class accuracy for COVID-19 was 98% with a 1.3 mis-classification rate as well as 100% precision and specificity; model was able to correctly classify 95 out of 100 COVID-19 images (Figure 8), a sensitivity rate of 95%. However, further improvement is required as the model mis-classified 1 image as Normal and 4 images as Viral pneumonia, as shown in Figure 8. The Normal and Viral Pneumonia classes also showed good performance with 93% and 92% accuracy, respectively. The sensitivity of the Normal class is 84%, which could be because of poor image acquisition process and needs to be improved on. The model trained efficiently with the training process showing no overfitting as the validation accuracies were better than the training accuracies as shown in Figure 2.

Furthermore, the result for the COVID-19 and Normal 2-class model as presented in Table 5 showed excellent performance with an overall accuracy and misclassification rate of 99.6% and 0.4% respectively. The COVID-19 class has 100% precision and specificity as well as 99% sensitivity as shown in Figure 9 with 1 image out of 100 COVID-19 images misclassified. The Normal class also performed well with no misclassification at 100% sensitivity as well as 99% specificity and precision, respectively. The model trained efficiently with no overfitting recorded during the model's training process as shown in Figure 4.

Lastly, the result for the COVID-19 and Viral Pneumonia 2-class model as presented in Table 6 demonstrated good performance with an overall accuracy and misclassification rate of 98.4% and 1.6% respectively. The COVID-19 class has precision and specificity of 100% respectively with 96% sensitivity. From Figure 10, the model misclassified 4 COVID-19 images as Viral Pneumonia, which could be due to similarities of the two infections. All the same, there was no misclassification in the Viral Pneumonia class, with 97%, 100%, and 96% precision, sensitivity, and specificity, respectively. The model's training process did not record any overfitting as shown in Figure 5 and the three models all show low false predictions (False negatives) which is a promising indication for medical application of these models with some improvements to further reduce false predictions.

## 5.1 Comparisons with reported heavyweight Models

### 5.1.1 3-class model

Table 7 presents performance comparison of MobileNetV2 used in this study and existing state-of-art that used heavyweight based deep learning models for 3-class (COVID-19, Normal and Viral Pneumonia) detection. The comparison is based on the number of parameters, memory size, accuracy, precision, recall, and F1-score.

**Table 7: Comparison of reported 3-class models**

| 3-Class Models | | | | | | | |
|---|---|---|---|---|---|---|---|
| Base model | Number of Parameters (Million) | Memory Usage (MB) | Dataset size | ACC (%) | PRE (%) | Recall (%) | F1-score (%) |
| Gupta *et al.* (2021) [53] **ResNet 101** | 44,654,504 | 171 | | 98 | 98 | 98.3 | 98 |
| **InceptionV3** | 24,000,000 | 92 | **COV**: 361 **NOR**: 365 **PNU**: 362 | 97 | 96.6 | 96.6 | 97 |
| **Xception** | 22,910,480 | 88 | | 97 | 97.3 | 97.3 | 97.6 |
| **InstaCOV-Net-19** (5 stacked CNN models) | 54,914,918 | - | | 99.08 | 99 | 99.3 | 99 |
| Heidari *et al.* (2020) [44] **VGG16** | 138,000,000 | 528 | **COV**: 415 **NOR**: 2880 **PNU**: 5179 | 94.5 | 88.3 | 95 | 91 |
| Wang *et al.* (2020) [22] **COVID-NET** | 11,750,000 | - | **COV**: 358 **NOR**: 8066 **PNU:** 5,538 | 93.3 | - | - | - |
| Zebin and Rezvy (2021) [41] **VGG16** | 138,000,000 | 528 | **COV**: 202 **NOR**: 300 **PNU**: 300 | 90 | 88 | - | - |
| **ResNet 50** | 26,000,000 | 99 | | 94.3 | 95.3 | - | - |
| **EfficientNetB0** | 5,300,000 | 29 | | 96.8 | 97.3 | - | - |
| Apostolopoulos *et al.* (2020) [25] **VGG19** | 143,667,240 | 549 | **COV:** 224 **NOR:** 504 **PNU:** 714 | 93.48 | 93 | 92.8 | 92.8 |
| **This Study** **MobileNetV2** | **3,538,984** | **14** | **COV: 1200 NOR: 1341 PNU: 1345** | 94.5 | 94.5 | 92.6 | 92.6 |

**Legend:** ACC= Accuracy, PRE= Precision, COV= COVID-19, NOR=Normal, PNU= Viral Pneumonia

The MobileNetV2 implementation yielded comparable performance to existing state-of-art that used heavyweight based deep learning models. Furthermore, it must be noted that MobileNetV2 is compute resource efficient as the models reported to date required between 6x to 15x greater number of parameters, as well as between 6x to 12x larger memory size inferring high computational, memory, and infrastructure costs. The training time for MobileNetV2 for the 3-class model is ~83 minutes, although the training times for the existing literature were not stated explicitly, the higher number of parameters used implies more computational time. In general, the performance of the lightweight model showed competitive performance to the heavyweight models with significant greater efficiency in terms of cost of computing resources such as memory size and number of parameters.

### 5.1.2   2-class models

Table 8 and Table 9 present a performance comparison between MobileNetV2 and existing state-of-art that used heavyweight based deep learning models for 2-class (COVID-19 and Normal) and (COVID-19 and Viral Pneumonia) COVID-19 detection. The former is more compute resource efficient requiring a training time of ~57 minutes, between 2.8x to 37x less memory as well as between 2x and 39x less parameters. MobileNetV2 implementation provides comparable performance and has the potential to be basis for developing low-cost mobile decision support applications.

**Table 8: Comparison of reported COVID-19 and Viral Pneumonia models**

| 2-Class Models (COVID-19 and Viral Pneumonia) | | | | | | | | |
|---|---|---|---|---|---|---|---|---|
| Base Model | Parameters (Millions) | Memory (MB) | Dataset size | ACC % | Recall | SPE % | PRE % | F1 % score |
| Narin *et al.* (2020) [54] **InceptionV3** | 24,000000 | 92 | COVID-19: **341** V. Pnu: **1493** | 98.6 | 99.7 | 98.3 | 93.2 | 96.3 |
| **ResNet 50** | 26,000000 | 98 | | 99.5 | 99.4 | 99.5 | 98.0 | 98.7 |
| **ResNet 101** | 44,654,504 | 171 | | 97.1 | 88.3 | 99.1 | 95.6 | 91.8 |
| **ResNet 152** | 60,344,232 | 232 | | 97.5 | 90.9 | 99.1 | 95.7 | 93.2 |
| **Inception ResNetV2** | 55,800000 | 215 | | 94.4 | 92.1 | 94.9 | 80.5 | 85.9 |
| **This study** **MobileNetV2** | **3,538,984** | **14** | COVID-19: **1200** V. Pnu: **1341** | **98.4** | 98 | 98 | 98.5 | 98.5 |

Table 9: Comparison of reported COVID-19 and Normal models

| 2-Class Models (COVID-19 and Normal) | | | | | | | | |
|---|---|---|---|---|---|---|---|---|
| Base Model | Parameters (Millions) | Memory (MB) | Dataset size | ACC % | Recall | SPE % | PRE % | F1 % score |
| Nayak *et al.* (2021) [42] ResNet 34 | 21,500000 | | COVID-19: **341** Normal: **2800** | 98.33 | - | 96.67 | 96.77 | 98 |
| GoogleNet | 7,000000 | 40 | | 96.67 | - | 96.67 | 96.67 | 96.67 |
| AlexNet | 60,000000 | 217 | | 97.50 | - | 96.67 | 96.72 | 97.52 |
| VGG16 | 138,000000 | 528 | | 95.83 | - | 95.0 | 95.08 | 95.87 |
| Gupta *et al.* (2021) [53] InceptionV3 | 24,000000 | 92 | COVID-19: **361** Normal: **365** | 97 | 97 | - | 97 | 97 |
| ResNet 101 | 44,654,504 | 171 | | 99 | 99 | - | 100 | 99 |
| Xception | 22,910,480 | 88 | | 99 | 99 | - | 100 | 99 |
| **This study** **MobileNetv2** | **3,538,984** | **14** | COVID-19:**1200** Normal: **1341** | **99.6** | **99.5** | **99.5** | **99.5** | **99.5** |

**Legend:** ACC= Accuracy, SPE= specificity, PRE= Precision

## 6.0 Conclusions, Recommendations and Future Work

Most recent research in the use of deep learning in COVID-19 detection using chest X-ray images has focused on heavyweight CNN models that required substantial computing power and therefore not amenable to deployment on resource constrained devices such as smartphones, which have enjoyed ubiquitous adoption in recent times. Mobile on-device medical imaging systems for COVID-19 detection are being advocated as a crucial tool to satisfy the prohibitive demand for rapid diagnosis, central to curb the spread of the pandemic, especially in rural and economically challenged areas.

The demand has motivated the evaluation of lightweight deep-CNN models aligned with the implementation requirements for use in mobile devices for efficient COVID-19 diagnosis using chest X-ray images. Here, transfer learning and data augmentation methods were applied to train MobileNetV2, a lightweight deep learning model for COVID-19 detection. The development and evaluation of performance used a chest X-ray dataset with 3200 images. Three models were developed and tested; a 3-class model, which classified the chest X-ray images into COVID-19, Normal and Viral Pneumonia; and two 2-class models, which classified the chest X-ray images into (COVID-19 and Normal) and (COVID-19 and Viral Pneumonia) respectively. The models were evaluated on a blind test set (i.e., not part of the data used for training) for unbiased assessment using standard statistical metrics. Results provide initial evidence that MobileNetv2 has the potential to detect COVID-19 using chest X-ray images with low false predictions (False Negatives) and be the basis for am impactful medical application.

Comparisons of performance with reported heavyweight deep-CNN implementations based on the number of parameters, memory size, accuracy, precision, recall, and F1-score showed that the performance of the lightweight was comparable to heavyweight models but at significantly greater efficiency in terms of cost of compute resources such as memory size and number of parameters. Although the initial evidence is that lightweight deep CNN algorithms have the potential of being used for COVID-19 detection, further improvements to the model are needed to reduce false predictions, a major limitation surfaced by the performance evaluation. However, sufficient evidence has been provided to further prove the feasibility of using lightweight deep learning models for COVID-19 detection delivered in a low cost, rapid, mobile automated decision support application. Once proven in operational environments, further benefits can potentially result from the diagnosis of other chest-related illness conditions such as tuberculosis.

Future model developments should focus reducing the false predictions (false negatives) and increasing the sensitivity as more COVID-19 images becomes available. Furthermore, the selectivity of the classification of COVID-19 from other SARS viruses remains to be proven as is the implementation and test of the model as a mobile application in operational environments

**Conflict of Interest**
The authors declare no conflicts of interest.

# REFERENCES

[1] W. Herring, *Learning Radiology: Recognizing the Basics*. Elsevier Health Sciences, 2007.
[2] 'WHO | Medical imaging', *WHO*. http://www.who.int/diagnostic_imaging/en/ (accessed Mar. 06, 2021).
[3] J. Beutel, H. L. Kundel, and R. L. Van Metter, *Handbook of medical imaging*, vol. 1. Spie Press, 2000.
[4] 'COVID-19 cases worldwide by day', *Statista*. https://www.statista.com/statistics/1103040/cumulative-coronavirus-covid19-cases-number-worldwide-by-day/ (accessed Apr. 23, 2022).
[5] 'Coronavirus deaths worldwide by country', *Statista*. https://www.statista.com/statistics/1093256/novel-coronavirus-2019ncov-deaths-worldwide-by-country/ (accessed Apr. 23, 2022).
[6] Y. Fang *et al.*, 'Sensitivity of Chest CT for COVID-19: Comparison to RT-PCR', *Radiology*, Feb. 2020, doi: 10.1148/radiol.2020200432.
[7] P. Huang *et al.*, 'Use of Chest CT in Combination with Negative RT-PCR Assay for the 2019 Novel Coronavirus but High Clinical Suspicion', *Radiology*, vol. 295, no. 1, pp. 22–23, Apr. 2020, doi: 10.1148/radiol.2020200330.
[8] H. Y. F. Wong *et al.*, 'Frequency and Distribution of Chest Radiographic Findings in Patients Positive for COVID-19', *Radiology*, vol. 296, no. 2, pp. E72–E78, Mar. 2020, doi: 10.1148/radiol.2020201160.
[9] W. Wang *et al.*, 'Detection of SARS-CoV-2 in Different Types of Clinical Specimens', *JAMA*, vol. 323, no. 18, pp. 1843–1844, May 2020, doi: 10.1001/jama.2020.3786.
[10] Z. Zhang and E. Sejdić, 'Radiological images and machine learning: Trends, perspectives, and prospects', *Computers in Biology and Medicine*, vol. 108, pp. 354–370, May 2019, doi: 10.1016/j.compbiomed.2019.02.017.

[11] M. Alzubaidi, H. D. Zubaydi, A. A. Bin-Salem, A. A. Abd-Alrazaq, A. Ahmed, and M. Househ, 'Role of deep learning in early detection of COVID-19: Scoping review', *Computer methods and programs in biomedicine update*, vol. 1, p. 100025, 2021.

[12] 'COVID-19 Radiography Database'. https://kaggle.com/tawsifurrahman/covid19-radiography-database (accessed Mar. 17, 2021).

[13] M. L. Giger, 'Machine Learning in Medical Imaging', *Journal of the American College of Radiology*, vol. 15, no. 3, Part B, pp. 512–520, Mar. 2018, doi: 10.1016/j.jacr.2017.12.028.

[14] W. Zhang, K. Doi, M. L. Giger, Y. Wu, R. M. Nishikawa, and R. A. Schmidt, 'Computerized detection of clustered microcalcifications in digital mammograms using a shift-invariant artificial neural network', *Medical Physics*, vol. 21, no. 4, pp. 517–524, 1994, doi: https://doi.org/10.1118/1.597177.

[15] M. H. Yap *et al.*, 'End-to-end breast ultrasound lesions recognition with a deep learning approach', in *Medical Imaging 2018: Biomedical Applications in Molecular, Structural, and Functional Imaging*, Mar. 2018, vol. 10578, p. 1057819. doi: 10.1117/12.2293498.

[16] Al-Antari MA, Al-Masni MA, Park SU, Park J, Metwally MK, Kadah YM, Han SM, Kim TS. An automatic computer-aided diagnosis system for breast cancer in digital mammograms via deep belief network. Journal of Medical and Biological Engineering. 2018 Jun;38(3):443-56.

[17] M. Byra *et al.*, 'Breast mass classification in sonography with transfer learning using a deep convolutional neural network and color conversion', *Medical Physics*, vol. 46, no. 2, pp. 746–755, 2019, doi: https://doi.org/10.1002/mp.13361.

[18] T. Fujioka *et al.*, 'Distinction between benign and malignant breast masses at breast ultrasound using deep learning method with convolutional neural network', *Jpn J Radiol*, vol. 37, no. 6, pp. 466–472, Jun. 2019, doi: 10.1007/s11604-019-00831-5.

[19] 'An enhanced deep learning approach for brain cancer MRI images classification using residual networks', *Artificial Intelligence in Medicine*, vol. 102, p. 101779, Jan. 2020, doi: 10.1016/j.artmed.2019.101779.

[20] A. Rehman, S. Naz, M. I. Razzak, F. Akram, and M. Imran, 'A Deep Learning-Based Framework for Automatic Brain Tumors Classification Using Transfer Learning', *Circuits Syst Signal Process*, vol. 39, no. 2, pp. 757–775, Feb. 2020, doi: 10.1007/s00034-019-01246-3.

[21] A. M. Sarhan, 'Brain Tumor Classification in Magnetic Resonance Images Using Deep Learning and Wavelet Transform', *Journal of Biomedical Science and Engineering*, vol. 13, no. 06, Art. no. 06, Jun. 2020, doi: 10.4236/jbise.2020.136010.

[22] L. Wang, Z. Q. Lin, and A. Wong, 'Covid-net: A tailored deep convolutional neural network design for detection of covid-19 cases from chest x-ray images', *Scientific Reports*, vol. 10, no. 1, pp. 1–12, 2020.

[23] B. Abraham and M. S. Nair, 'Computer-aided detection of COVID-19 from X-ray images using multi-CNN and Bayesnet classifier', *Biocybernetics and Biomedical Engineering*, vol. 40, no. 4, pp. 1436–1445, Oct. 2020, doi: 10.1016/j.bbe.2020.08.005.

[24] N. Wang, H. Liu, and C. Xu, 'Deep Learning for The Detection of COVID-19 Using Transfer Learning and Model Integration', in *2020 IEEE 10th International Conference on Electronics Information and Emergency Communication (ICEIEC)*, Jul. 2020, pp. 281–284. doi: 10.1109/ICEIEC49280.2020.9152329.

[25] I. D. Apostolopoulos and T. A. Mpesiana, 'Covid-19: automatic detection from X-ray images utilizing transfer learning with convolutional neural networks', *Phys Eng Sci Med*, vol. 43, no. 2, pp. 635–640, Jun. 2020, doi: 10.1007/s13246-020-00865-4.

[26] L. O. Hall, R. Paul, D. B. Goldgof, and G. M. Goldgof, 'Finding covid-19 from chest x-rays using deep learning on a small dataset', *arXiv preprint arXiv:2004.02060*, 2020.

[27] H. S. Maghdid, A. T. Asaad, K. Z. Ghafoor, A. S. Sadiq, and M. K. Khan, 'Diagnosing COVID-19 Pneumonia from X-Ray and CT Images using Deep Learning and Transfer Learning Algorithms', p. 8.

[28] A. I. Khan, J. L. Shah, and M. M. Bhat, 'CoroNet: A deep neural network for detection and diagnosis of COVID-19 from chest x-ray images', *Computer Methods and Programs in Biomedicine*, vol. 196, p. 105581, Nov. 2020, doi: 10.1016/j.cmpb.2020.105581.

[29] F. Ucar and D. Korkmaz, 'COVIDiagnosis-Net: Deep Bayes-SqueezeNet based diagnosis of the coronavirus disease 2019 (COVID-19) from X-ray images', *Medical Hypotheses*, vol. 140, p. 109761, Jul. 2020, doi: 10.1016/j.mehy.2020.109761.

[30] T. Ozturk, M. Talo, E. A. Yildirim, U. B. Baloglu, O. Yildirim, and U. Rajendra Acharya, 'Automated detection of COVID-19 cases using deep neural networks with X-ray images', *Computers in Biology and Medicine*, vol. 121, p. 103792, Jun. 2020, doi: 10.1016/j.compbiomed.2020.103792.

[31] H. Mukherjee, S. Ghosh, A. Dhar, S. M. Obaidullah, K. C. Santosh, and K. Roy, 'Deep neural network to detect COVID-19: one architecture for both CT Scans and Chest X-rays', *Appl Intell (Dordr)*, vol. 51, no. 5, pp. 2777–2789, 2021, doi: 10.1007/s10489-020-01943-6.

[32] W. Zouch *et al.*, 'Detection of COVID-19 from CT and Chest X-ray Images Using Deep Learning Models', *Ann Biomed Eng*, Apr. 2022, doi: 10.1007/s10439-022-02958-5.

[33] A. Bhattacharyya, D. Bhaik, S. Kumar, P. Thakur, R. Sharma, and R. B. Pachori, 'A deep learning based approach for automatic detection of COVID-19 cases using chest X-ray images', *Biomedical Signal Processing and Control*, vol. 71, p. 103182, Jan. 2022, doi: 10.1016/j.bspc.2021.103182.

[34] J. Koh, Y. Yoon, S. Kim, K. Han, and E.-K. Kim, 'Deep Learning for the Detection of Breast Cancers on Chest Computed Tomography', *Clinical Breast Cancer*, vol. 22, no. 1, pp. 26–31, Jan. 2022, doi: 10.1016/j.clbc.2021.04.015.

[35] M. Soni, S. Gomathi, P. Kumar, P. P. Churi, M. A. Mohammed, and A. O. Salman, 'Hybridizing Convolutional Neural Network for Classification of Lung Diseases', *IJSIR*, vol. 13, no. 2, pp. 1–15, Apr. 2022, doi: 10.4018/IJSIR.287544.

[36] A. M. Alqudah, S. Qazan, and A. Alqudah, 'Automated Systems for Detection of COVID-19 Using Chest X-ray Images and Lightweight Convolutional Neural Networks', In Review, preprint, Apr. 2020. doi: 10.21203/rs.3.rs-24305/v1.

[37] M. Chakraborty, S. V. Dhavale, and J. Ingole, 'Corona-Nidaan: lightweight deep convolutional neural network for chest X-Ray based COVID-19 infection detection', *Appl Intell*, vol. 51, no. 5, pp. 3026–3043, May 2021, doi: 10.1007/s10489-020-01978-9.

[38] S. Karakanis and G. Leontidis, 'Lightweight deep learning models for detecting COVID-19 from chest X-ray images', *Computers in Biology and Medicine*, vol. 130, p. 104181, Mar. 2021, doi: 10.1016/j.compbiomed.2020.104181.

[39] S. Bekhet, M. H. Alkinani, R. Tabares-Soto, and M. Hassaballah, 'An efficient method for covid-19 detection using light weight convolutional neural network', *Computers, Materials and Continua*, pp. 2475–2491, 2021.

[40] M.-L. Huang and Y.-C. Liao, 'A lightweight CNN-based network on COVID-19 detection using X-ray and CT images', *Computers in Biology and Medicine*, vol. 146, p. 105604, Jul. 2022, doi: 10.1016/j.compbiomed.2022.105604.

[41] T. Zebin and S. Rezvy, 'COVID-19 detection and disease progression visualization: Deep learning on chest X-rays for classification and coarse localization', *Appl Intell*, vol. 51, no. 2, pp. 1010–1021, Feb. 2021, doi: 10.1007/s10489-020-01867-1.

[42] S. R. Nayak, D. R. Nayak, U. Sinha, V. Arora, and R. B. Pachori, 'Application of deep learning techniques for detection of COVID-19 cases using chest X-ray images: A comprehensive study', *Biomedical Signal Processing and Control*, vol. 64, p. 102365, Feb. 2021, doi: 10.1016/j.bspc.2020.102365.

[43] T. Sanida, A. Sideris, D. Tsiktsiris, and M. Dasygenis, 'Lightweight neural network for COVID-19 detection from chest X-ray images implemented on an embedded system', *Technologies*, vol. 10, no. 2, p. 37, 2022.

[44] M. Heidari, S. Mirniaharikandehei, A. Z. Khuzani, G. Danala, Y. Qiu, and B. Zheng, 'Improving the performance of CNN to predict the likelihood of COVID-19 using chest X-ray

images with preprocessing algorithms', *International Journal of Medical Informatics*, vol. 144, p. 104284, Dec. 2020, doi: 10.1016/j.ijmedinf.2020.104284.

[45] S. Asif, M. Zhao, F. Tang, and Y. Zhu, 'A deep learning-based framework for detecting COVID-19 patients using chest X-rays', *Multimedia Systems*, pp. 1–19, 2022.

[46] S. V. Kogilavani *et al.*, 'COVID-19 detection based on lung CT scan using deep learning techniques', *Computational and Mathematical Methods in Medicine*, vol. 2022, 2022.

[47] N. Rajawat, B. S. Hada, M. Meghawat, S. Lalwani, and R. Kumar, 'C-COVIDNet: A CNN Model for COVID-19 Detection Using Image Processing', *Arabian Journal for Science and Engineering*, pp. 1–12, 2022.

[48] N. Subramanian, O. Elharrouss, S. Al-Maadeed, and M. Chowdhury, 'A review of deep learning-based detection methods for COVID-19', *Computers in Biology and Medicine*, p. 105233, 2022.

[49] S. J. Pan and Q. Yang, 'A Survey on Transfer Learning', *IEEE Transactions on Knowledge and Data Engineering*, vol. 22, no. 10, pp. 1345–1359, Oct. 2010, doi: 10.1109/TKDE.2009.191.

[50] O. Russakovsky *et al.*, 'ImageNet Large Scale Visual Recognition Challenge', *arXiv:1409.0575 [cs]*, Jan. 2015, Accessed: Mar. 22, 2021. [Online]. Available: http://arxiv.org/abs/1409.0575

[51] M. Sandler, A. Howard, M. Zhu, A. Zhmoginov, and L.-C. Chen, 'MobileNetV2: Inverted Residuals and Linear Bottlenecks', *arXiv:1801.04381 [cs]*, Mar. 2019, Accessed: Mar. 17, 2021. [Online]. Available: http://arxiv.org/abs/1801.04381

[52] M. Hossin and M. N. Sulaiman, 'A review on evaluation metrics for data classification evaluations', *International journal of data mining & knowledge management process*, vol. 5, no. 2, p. 1, 2015.

[53] A. Gupta, Anjum, S. Gupta, and R. Katarya, 'InstaCovNet-19: A deep learning classification model for the detection of COVID-19 patients using Chest X-ray', *Applied Soft Computing*, vol. 99, p. 106859, Feb. 2021, doi: 10.1016/j.asoc.2020.106859.

[54] A. Narin, C. Kaya, and Z. Pamuk, 'Automatic Detection of Coronavirus Disease (COVID-19) Using X-ray Images and Deep Convolutional Neural Networks', *arXiv:2003.10849 [cs, eess]*, Oct. 2020, Accessed: Apr. 23, 2021. [Online]. Available: http://arxiv.org/abs/2003.10849